\documentclass[12pt,preprint]{aastex} \usepackage{graphicx}

\shorttitle{Multiple Populations in NGC 6752}
\shortauthors{Milone et al.}
\usepackage{ulem}

\begin{document}

\title{Multiple stellar populations in the Galactic globular cluster NGC 6752
          \footnote{           Based on observations with  the
                               NASA/ESA {\it Hubble Space Telescope},
                               obtained at  the Space Telescope Science
                               Institute,  which is operated by AURA, Inc.,
                               under NASA contract NAS 5-26555.}}

\author{
A.\ P. Milone\altaffilmark{2},
G.\ Piotto\altaffilmark{2},
I.\ R.\ King\altaffilmark{3},
L.\ R.\ Bedin\altaffilmark{4},
J.\ Anderson\altaffilmark{4},
A.\ F. Marino\altaffilmark{2,5},
Y.\ Momany\altaffilmark{6},
L.\ Malavolta\altaffilmark{2,7},
and
S.\ Villanova\altaffilmark{8}
}

\altaffiltext{2}{Dipartimento di Astronomia, Universit\`a di Padova, Vic. Osservatorio 3,
 I-35122, PD, Italy;
 [antonino.milone,giampaolo.piotto,anna.marino,luca.malavolta]@unipd.it}

\altaffiltext{3}{Department  of  Astronomy,  University  of  Washington,
  Seattle, WA 98195-1580, USA; king@astro.washington.edu}

\altaffiltext{4}{Space Telescope Science Institute, 3800 San Martin
 Drive, Baltimore, MD 21218; [jayander,bedin]@stsci.edu}

\altaffiltext{5}{P. Universidad Cat\'olica de Chile, Departamento de
 Astronom\'ia y Astrof\'isica, Casilla 306, Santiago 22, Chile; fmarino@astro.puc.cl}

\altaffiltext{6}{Osservatorio Astronomico di Padova, Vicolo
 dell'Osservatorio 5, 35122 Padova, Italy; yazan.momany@oapd.inaf.it}

\altaffiltext{7}{Visiting undergraduate Student at STScI under the
   {\it 2009 Space Astronomy Summer Program},
   {\sf http://www.stsci.edu/institute/sd/students}}

\altaffiltext{8}{Departamento de Astronomia, Universidad de Concepcion,
  Casilla 160-C, Concepcion, Chile; svillanova@astro-udec.cl}

\begin{abstract}
We have carried out high-precision photometry on a large number of
archival {\it HST\/} images of the Galactic globular cluster NGC 6752, to
search for signs of multiple stellar populations.  We find a broadened
main sequence, and demonstrate that this broadening cannot be attributed
either to binaries or to photometric errors.  There is also some
indication of a main-sequence split.
No significant spread could be found along the subgiant branch,
however.
Ground-based photometry reveals that in the $U$ vs.\ $(U-B)$
color-magnitude diagram the red-giant branch exhibits a clear color
spread, which we have been able to correlate with variations in Na and O
abundances.  In particular the Na-rich, O-poor stars identified by
Carretta et al.\ (2007) define a sequence on the red side of the
red-giant branch, while Na-poor, O-rich stars populate a bluer, more
dispersed portion of the red-giant branch.
\end{abstract}


\section{Introduction}
\label{introduction}

For several decades astronomers have believed that globular clusters
(GCs) were made up of stars that were all born at the same time and out
of the same material, and that as such, they offered the best
approximation to a simple stellar population (see, e.g., Renzini \&
Buzzoni 1986).
The abundance anomalies observed among GC stars since the early
seventies (see Gratton et al.\ 2004 for a recent review) have challenged
this scenario, but it was only the recent discovery of clear photometric
evidence of multiple stellar populations among unevolved stars that
destroyed once and for all this traditional scenario, moving the study
of GCs in a new direction.

The most striking case is $\omega$ Centauri, which exhibits large
star-to-star iron variations (see Villanova et al.\ 2007 and references
therein) with evidence of multiple red-giant branches (RGBs, seen earlier
by Lee et al.\ 1999 and Pancino et al.\ 2000), multiple subgiant
branches (SGBs), and at least three distinct main sequences (MSs, Bedin et
al.\ 2004).  Furthermore, Piotto et al.\ (2005) showed that the red MS
is more metal poor than the blue MS, which may imply that the latter has
a strong He enhancement (as suggested by Norris 2004).

High-accuracy photometry from {\it Hubble Space Telescope} ({\it HST})
images has allowed us to discover that NGC 2808 also hosts multiple
distinct MSs (in this case three; see Piotto et al.\ 2007), possibly
associated with three stellar populations with different He content and
with the multi-modal horizontal branch (HB, D'Antona et al.\ 2005).  In
addition, we have detected a split SGB in NGC 1851 (Milone et al.\
2008), M22 (NGC 6656; Piotto\ 2009, Marino et al.\ 2009), NGC 6388
(Moretti et al.\ 2009), and many other GCs, both in the Milky Way
(Piotto 2009) and in the Magellanic Clouds (Mackey et al.\ 2008, Milone
et al.\ 2009).  More recently, Anderson et al.\ (2009) have found that
the MS of 47 Tuc (NGC 104) is also spread much more than can be expected
from photometric errors, while the SGB shows at least two distinct
components.

The nearby globular cluster NGC 6752 ($d=4.0$ kpc, $M_V=-7.73$, Harris
1996) is an\-other very promising candidate to examine for multiple
populations.  There is evidence of strong abundance anomalies (Carretta
et al.\ 2007), and it has an extended blue HB, similar to the HBs of NGC
2808, $\omega$ Cen, and M54 (NGC 6715), all clusters in which mixed
stellar populations have recently been found.

In this paper we will use both {\it HST} and ground-based photometry to
study the color-magnitude diagram (CMD) of NGC 6752 for signs of
multiple stellar populations.  {\it HST} data will allow us to carefully
analyze the MS and the SGB in the central field, while ground-based data
will be used to examine the RGB.  One careful photometric study of the
MS of NGC 6752 already exists:\ Rubenstein \& Bailyn (1997, RB97) used
{\sl HST\/}'s WFPC2 camera to study the MS of the cluster, and concluded
that there was a broadening toward the red side that could be explained
by the presence of a considerable fraction of binaries near the center
of the cluster.  We will reconcile our results with those of RB97, by
re-reducing their images with the techniques of today and by introducing
new measurements of images from {\sl HST\/}'s Advanced Camera for
Surveys (ACS), as well as ground-based images with a larger field.

\newpage
\section{Observations and data reduction}
\label{data}
%

\begin{figure}
   \epsscale{.80}
   \plotone{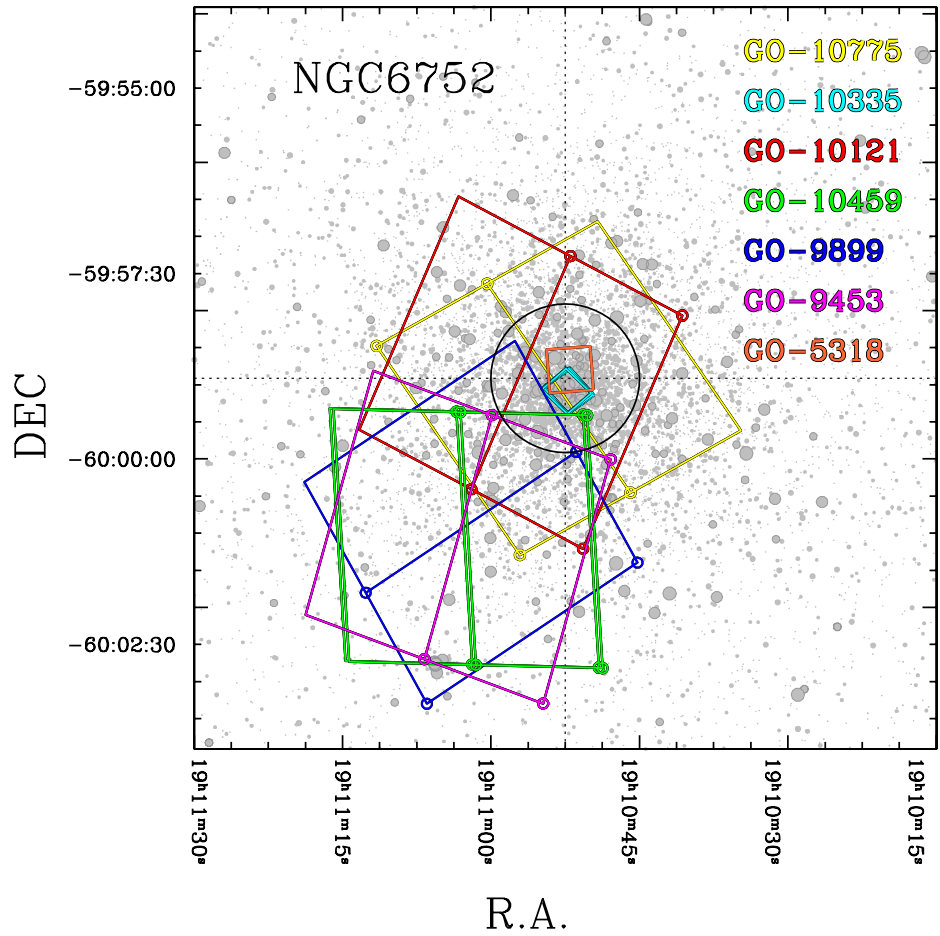}
   \caption{Footprints of the {\it HST\/}/ACS data sets used in this
   work.
   The small orange box at the center is the footprint of the PC
   field of R\&B.
   The small circles mark the corners of
   WFC chip \#1.}
   \label{footprint}
   \end{figure}

In order to search for signs of multiple stellar populations in the
crowded core of NGC 6752, we retrieved a large and varied set of images
from the {\it HST} archive.  These images are listed in Table 1, while
Fig.~\ref{footprint} shows their footprints.  One of the problems of
identifying multiple populations, or else spreads in cluster sequences,
is the fact that photometric errors can introduce similar signatures.
Anderson et al.\ (2009) have shown that a good way to distinguish real
broadening from mere photometric error is to analyze independent data
sets, and see if they exhibit the same features.  Although the archival
material available for NGC 6752 it is not as extensive as the set of
images that Anderson et al.\ had for 47 Tuc, we were able to make use of
several different {\sl HST} programs whose ACS images used both the WFC
and the HRC channels, and were at different pointings and roll angles
and used different filter pass-bands.

From the archival {\sl HST} images of NGC 6752 in various passbands, we
were able to make five completely independent CMDs.  Four of our ACS
CMDs come from WFC images, and a fifth from HRC images:\\
1) $m_{\rm F555W}$ and $m_{\rm F814W}$ from GO-10121;\\
2) $m_{\rm F606W}$ and $m_{\rm F814W}$ from GO-10775;\\
3) $m_{\rm F475W}$ from GO-9899 and $m_{\rm F606W}$ from GO-10459;\\
4) $m_{\rm F606W}$ and $m_{\rm F814W}$ from GO-9453;\\
5) $m_{\rm F435W}$ and $m_{\rm F555W}$ from GO-10335 (both taken with
the HRC).\\
Data sets 1--4 allowed us to study the MS (Sect.\ \ref{MS}), while data
sets 1, 2, 4, and 5 were used to study the structure of the SGB (Sect.\
\ref{SGB}).

We also used, for comparison purposes, the WFPC2 images of Rubenstein \&
Bailyn (1997), from their GO-5318.  The results of our remeasurement of
those images will be presented in the following section.

The ACS/WFC images were reduced by using the procedure described in
Anderson et al.\ (2008), which allowed us to analyze all the exposures
of each data set simultaneously to generate a single star list.  Stars
are measured independently in each image by using a spatially varying
9$\times$10 array of empirical ``library PSFs'' from Anderson \& King
(2006), plus a spatially constant perturbation for each exposure, to
allow for variations in the telescope focus.  The software is able to
detect almost every star that would be found by eye. It was designed to
work well in both crowded and uncrowded fields, and takes advantage of
the many independent dithered pointings and the knowledge of the PSF, to
avoid including artifacts in the list.  The photometry was put into the
ACS Vega-mag system following recipes in Bedin et al.\ (2005) and using
the zero points given in Sirianni et al.\ (2005).  Unfortunately, the
hybrid PSF model above is not able to account for all of the effects of
telescope breathing, which can introduce a small spatial dependence of
the shape of the PSF, which is not compensated for in our PSF model and
can cause small systematic photometric errors that depend on position on
the detector.  The typical variation is small (about 1\% in the fraction
of light in the core).  To account for the color differences that these
variations produce, 
 we used the following procedure:\ first we drew a main-sequence ridge
line (MSRL), by putting a spline through the median colors found in
successive short intervals of magnitude, and we iterated this step with a
sigma clipping; then we examined the color residuals relative to this
sequence, as a function of location on the detector.  These color
variations can come from either differential reddening or from PSF
variation, but we make empirical star-by-star corrections, regardless of
the cause.  We compute for each star its color residual from the MSRL,
and then correct the star's color by the difference between its color
residual and the mean of those of its best-measured neighbors (50 to 100
of them, according to the detector and the star density).  These
corrections are typically smaller than 0.005 mag, and never exceed 0.013
mag, but including them makes a significant improvement in our
photometric results.

The measurement of stellar fluxes and positions in each ACS/HRC image
was performed by using the publicly available measuring routine, library
PSFs, and the distortion correction described in Anderson \& King
(2004).  We corrected the zero points of color in the same way as for
the WFC photometry.

As for the PC images from GO-5318, we used the same procedures as for
other WFPC2 images in Bedin et al.\ (2001).

Finally, since our focus here is on high-quality photometry, we included
in the analysis only relatively isolated, unsaturated stars with good
values of the PSF-fit quality index and small rms errors in photometry
and in astrometry.  A detailed description of the selection procedures
is given in Milone et al.\ (2009).

In addition, we used artificial-stars (AS) for several purposes: to
determine the completeness level of our sample,
to estimate the internal photometric errors, and to measure the fraction
of chance-superposition binaries.  The AS experiments followed
the recipes of Anderson et al.\ (2008), 
while the procedure used to determine a position-dependent completeness
is described in 
Milone et al.\ (2009).

\begin{table}
\center
\scriptsize {
\begin{tabular}{cccccl}
\hline
\hline
 INSTR &  DATE & N$\times$EXPTIME & FILTER  & PROGRAM & PI \\
\hline
WFC & Sep 15 2002 & 1$\times$4s   + 1$\times$40s    & F606W  & GO-9453  &  Brown      \\
WFC & Sep 15 2002 & 1$\times$4s   + 1$\times$46s    & F814W  & GO-9453  &  Brown      \\
WFC & Jul 18 2004 & 6$\times$340s                   & F475W  & GO-9899  &  Piotto     \\
WFC & Sep 19 2004 & 11$\times$435s + 12$\times$80s  & F555W  & GO-10121 &  Bailyn     \\
WFC & Sep 19 2004 & 12$\times$40s	            & F814W  & GO-10121 &  Bailyn     \\
WFC & Oct 16 2005 & 8$\times$450s                   & F606W  & GO-10459 &  Biretta    \\
WFC & May 24 2006 & 1$\times$2s   + 4$\times$35s    & F606W  & GO-10775 &  Sarajedini \\
WFC & May 24 2006 & 1$\times$2s   + 4$\times$40s    & F814W  & GO-10775 &  Sarajedini \\
HRC & Jun 08 2004 and Jun 05 2006 & 24$\times$35s   & F435W  & GO-10335 &  Ford     \\
HRC & Jun 08 2004 and Jun 05 2006 & 12$\times$10s   & F555W  & GO-10335 &  Ford    \\
WFPC2 & Aug 18-19 1994 &  117$\times$26s+ 26$\times$80s  & F555W  & GO-5318  &  Bailyn  \\
WFPC2 & Aug 18-19 1994 &  107$\times$50s+ 27$\times$160s & F814W  & GO-5318  &  Bailyn  \\
WFI@2.2m & Jul 25-26 2000 & 4$\times$30s   + 4$\times$150s & $U$ & 065.L-0561 & Piotto \\
WFI@2.2m & Jul 25-26 2000 & 3$\times$5s    + 4$\times$10s  & $B$ & 065.L-0561 & Piotto \\
\hline
\hline
\end{tabular}
}
\label{tabdata}
\caption{Description of the data sets used in this paper. }
\end{table}

%
%

The {\sl HST} fields are too small to include a statistically
significant number of giant stars, so in order to study the distribution
of the stars along the RGB (Sect.\ \ref{RGB}), we analyzed the
photometric catalogs obtained from the Wide Field Imager (WFI) of the
ESO/MPI 2.2m telescope, already presented in Momany et al.\ (2002).  The
ground-based $U$ and $B$ images of NGC 6752 were taken on July 25--26,
2000. The WFI camera consists of eight 2048$\times$4096 EEV CCDs, with a
total field of view of 34$\times$33 arcmin. The exposure times (30s and
150s in $U$, 5s and 10s in $B$) were chosen in order to sample both the
bright RGB and the upper MS stars. Weather conditions were photometric,
with good seeing (better than 0.8 arcsec FWHM for all images). Basic
reductions of the CCD mosaic were performed using the IRAF package
MSCRED (Valdes 1998), while stellar photometry was performed using the
DAOPHOT and ALLFRAME programs (Stetson 1994). Finally, instrumental
magnitudes were calibrated to the {\sl UBV} standard system by
observing, on each of the eight chips, a field of standard stars from
Landolt (1992), during the same nights as the NGC 6752 observations.

   \begin{figure}
   \epsscale{.80}
   \plotone{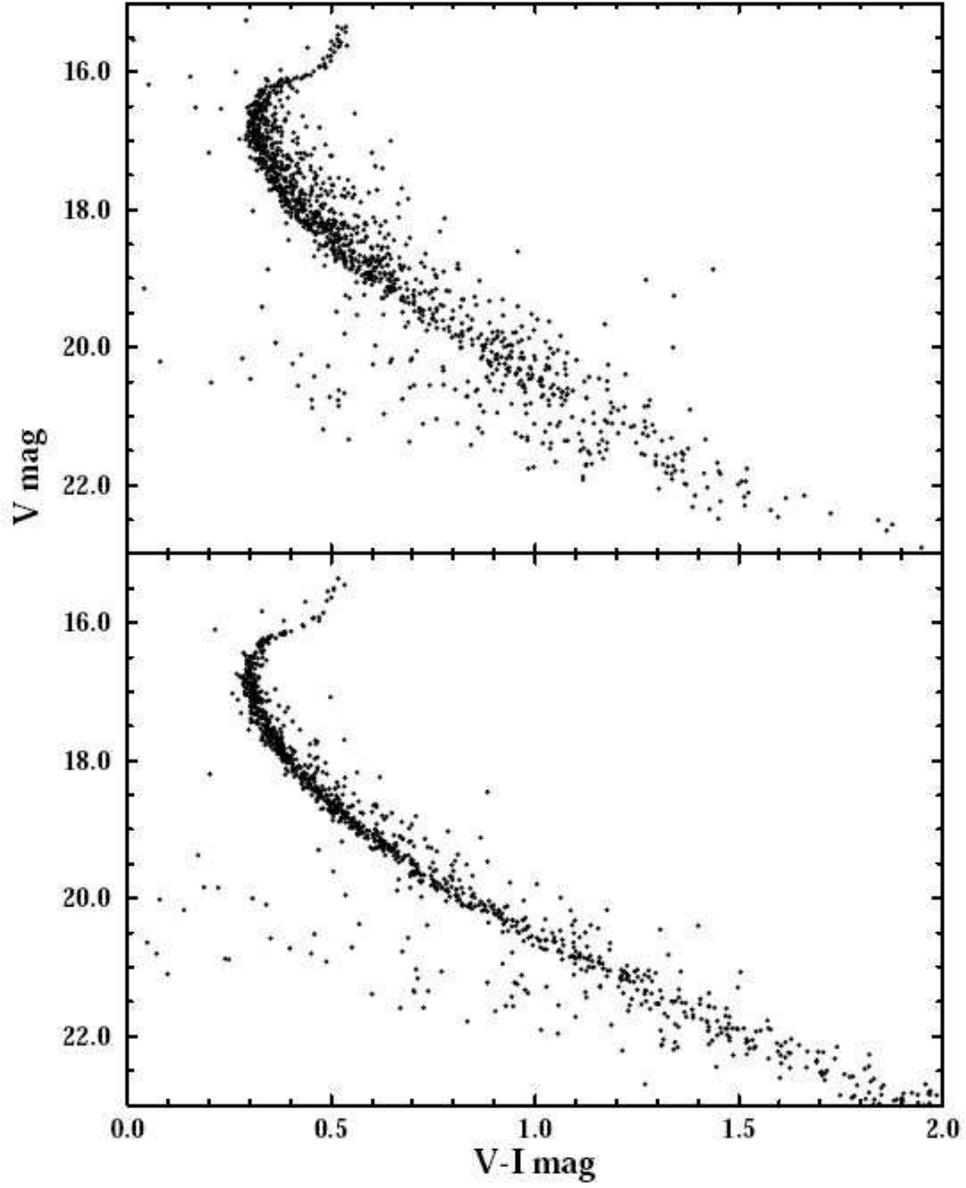}
   \caption{Reproduction of Fig.\ 5 of Rubenstein \& Bailyn (1997).  The
   upper half shows their CMD for the cluster core ($r<11$ arcsec) and
   the lower half their CMD for the remainder of the PC field.}
   \label{rb5}
   \end{figure}
%

\newpage
\section{The Main Sequence}
\label{MS}

Our aim here is to look for a possible broadening of the MS.  This could
come from an intrinsic spread in the character of the MS stars, but such
an appearance might also be the result of binaries, which are seen in
many clusters as a spread on the red side of the MS.  The organization
of this section is therefore as follows: First we take up the question
of binaries and their possible influence on our results, which we show
to be small in comparison with the broadening that we observe.  Then we
present the results from our best data set, from GO-10121, which show
the structure of the MS most clearly.  Finally, we confirm the MS spread
by comparing photometry from other ACS data sets with that from
GO-10121.

\subsection{The possible influence of binaries}

In a paper that set a standard of photometric accuracy for its time,
RB97 used {\sl HST\/}'s WFPC2 camera to study the MS of NGC 6752, and
concluded that there was a broadening toward the red side that could be
explained by the presence of an admixture of at least 15\% binaries near
the center of the cluster.  We now re-examine their result, first by
using the improved photometric techniques that are now available, and
then using the better data sets that have been collected more recently.

\begin{figure}
   \epsscale{.45} \plotone{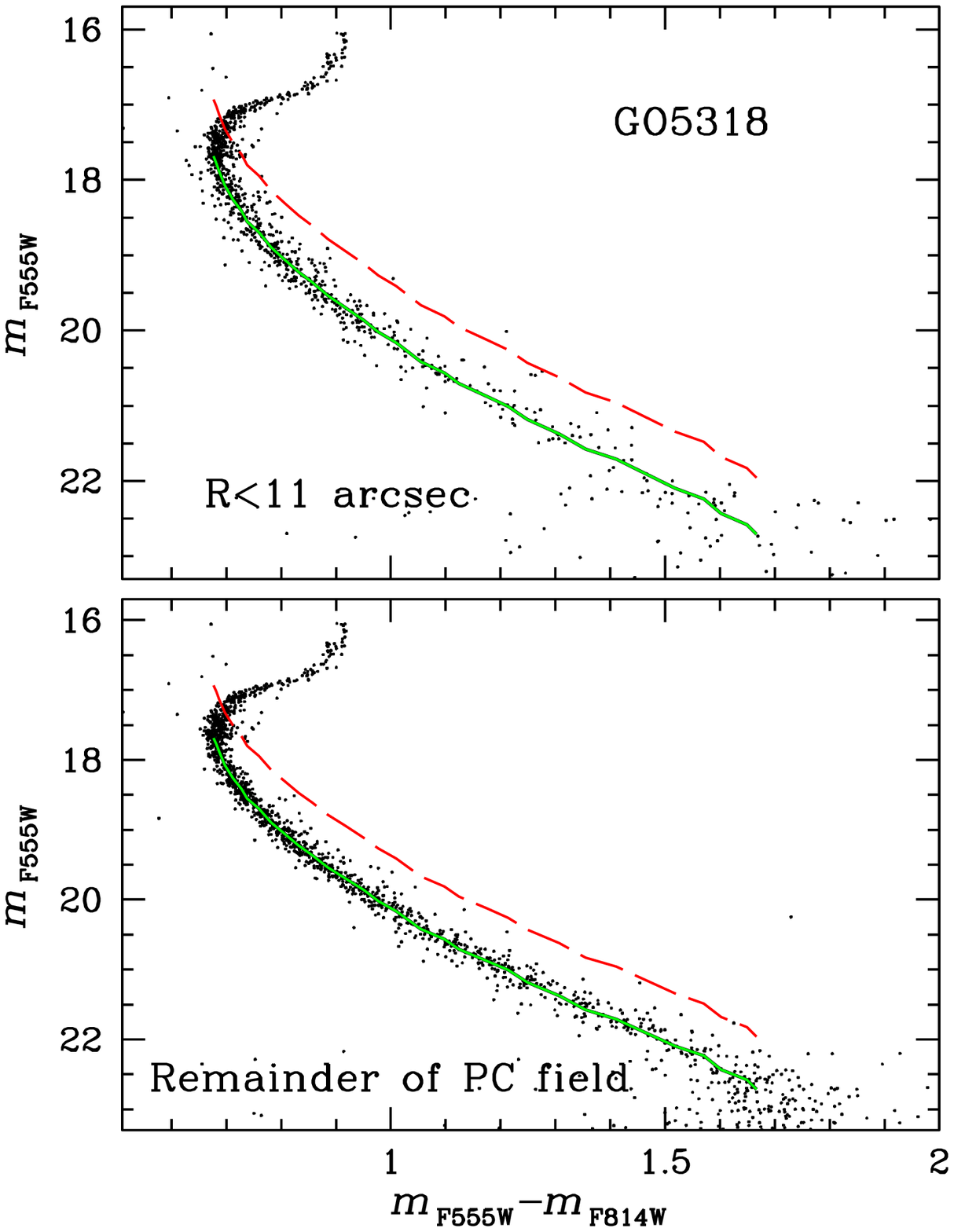} \plotone{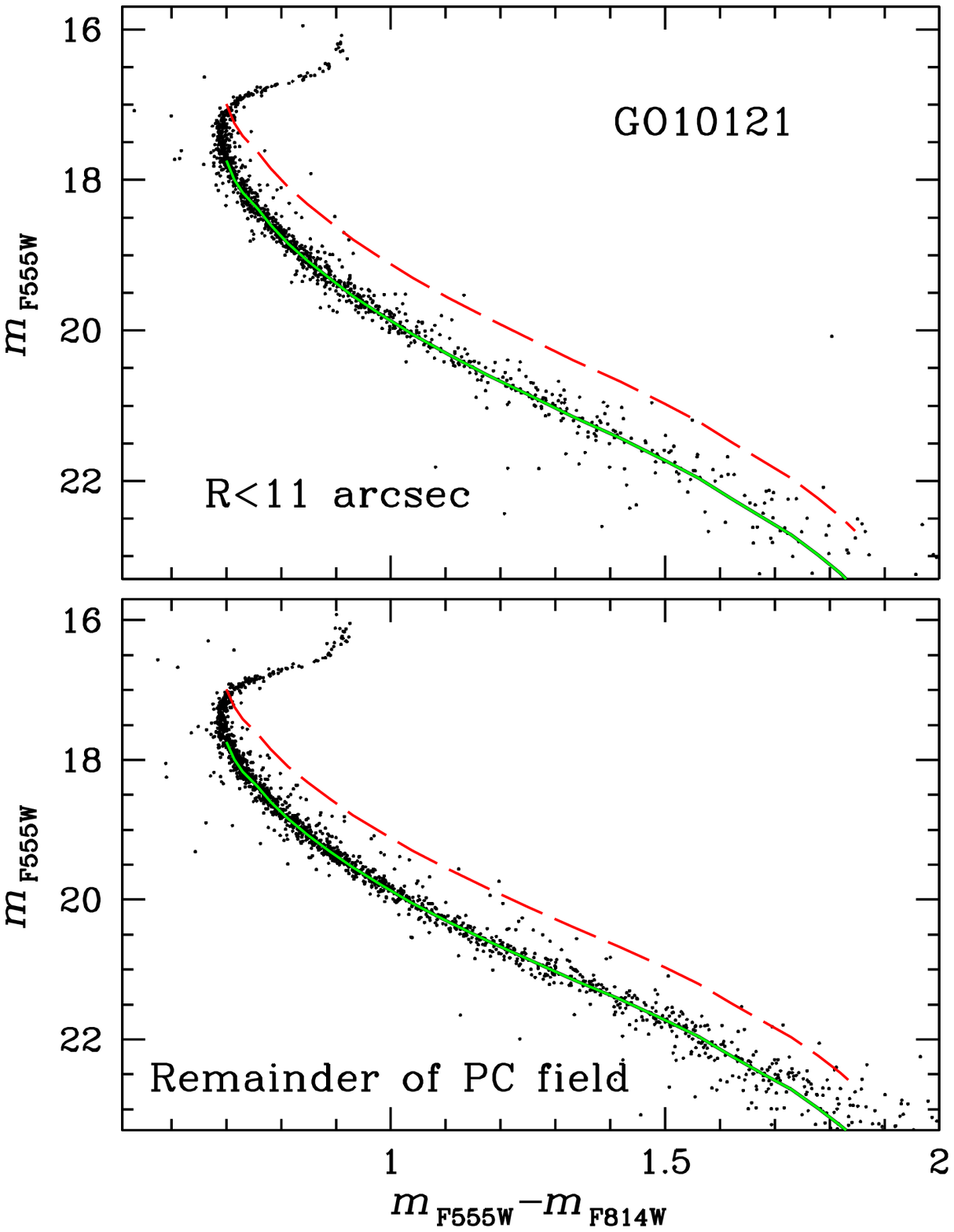}
   \caption{
   Left panels: As in Fig.\ \ref{rb5}, but showing our CMDs derived from
   the same images.  Right panels: CMDs for the same regions covered by
   the R\&B CMDs (Fig.\ \ref{rb5}) and the CMDs in the left panels, but
   from photometry done on the GO-10121 ACS images.  The dashed lines
   show the locus of equal-mass binaries.}
   \label{5318CMDs}
   \end{figure}

The data available to RB97, in program GO-5318 (P.I.\ Bailyn), consisted
of somewhat more than 100 PC images in each of the filters F555W and
F814W.  (Since their observations were made in a rapid-fire repetition,
limitations of downlink speed prevented them from recording the WF
data.)  Their measurements strongly suggested the presence of a quite
significant number of MS binaries near the cluster center, and a smaller
fraction in the outer part of their field, (as would be expected from
the greater central concentration of the binaries, on account of their
masses). Their Fig.\ 5 is reproduced here as our Fig.\ \ref{rb5}.  We
have remeasured their images, using the improved photometric techniques
of Anderson \& King (1999, 2000, 2003); our resulting CMDs are shown in
Fig.\ \ref{5318CMDs}, left panel.  The dashed lines in Fig.\
\ref{5318CMDs} show the locus of equal-mass binaries.  We now find, from
the images that RB97 used a dozen years ago, a strikingly smaller spread
than is stated in their paper.  The difference is due not to any error on
their part, but to the photometric improvement referred to above which
allows us better to resolve stellar blends.  We see that with our
improved measurements the evidence for binaries is almost completely
gone.  It is of course very unlikely that this cluster is devoid of
binaries; but they are concentrated to the central region, where they
represent only a tiny fraction of the cluster population.

%

The evidence for binaries is even further decreased when we use the
photometrically superior images that ACS provides (Fig.\
\ref{5318CMDs}, right panels).

      \begin{figure*}
      \epsscale{1} \plotone{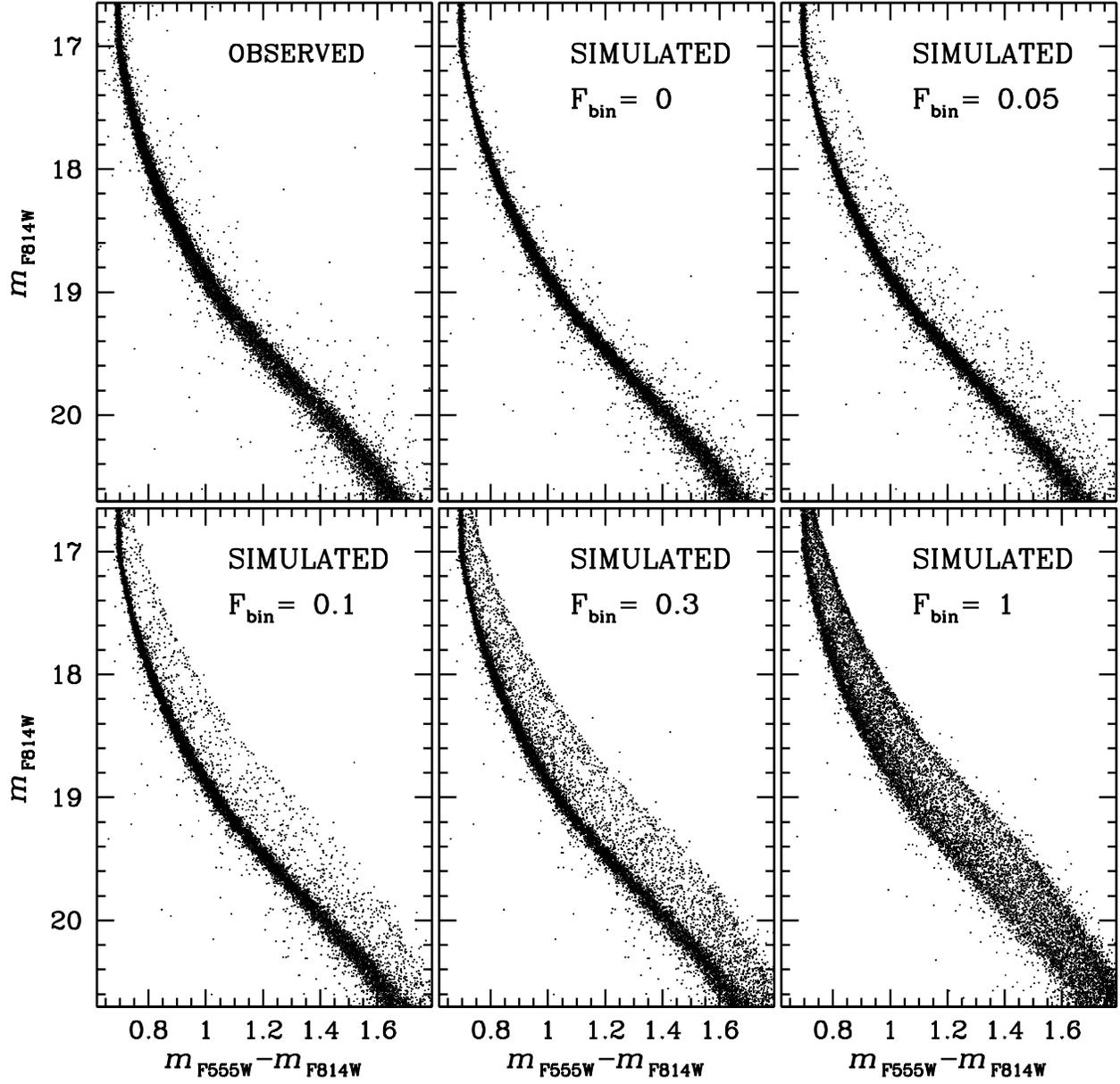}
   \caption{ Observed and simulated CMDs of NGC 6752. 
}
   \label{simulation}
   \end{figure*}
%

 In order to put a more quantitative constraint on the binary fraction in
NGC 6752, we created five simulated CMDs, assuming binary fractions
$F_{\rm bin}$ = 0.00, 0.05, 0.10, 0.30, 1.00, respectively.  
The simulations used random choices of artificial stars, as an
adaptation of the procedures of Sollima et al.\ (2007).  What we did was
 to make random choices from the Kroupa IMF (2002),
a star at a time for the MS, and also pairs to simulate binaries.  Using
a mass-luminosity relation (whose exact choice is unimportant) and an
assumed distance modulus, we calculated an $m_{\rm F814W}$ for each
star, and assigned the star the corresponding MS color; finally for each
binary we summed the flux in each band and derived the resulting
magnitude and color.

\begin{figure*}
\epsscale{1} \plotone{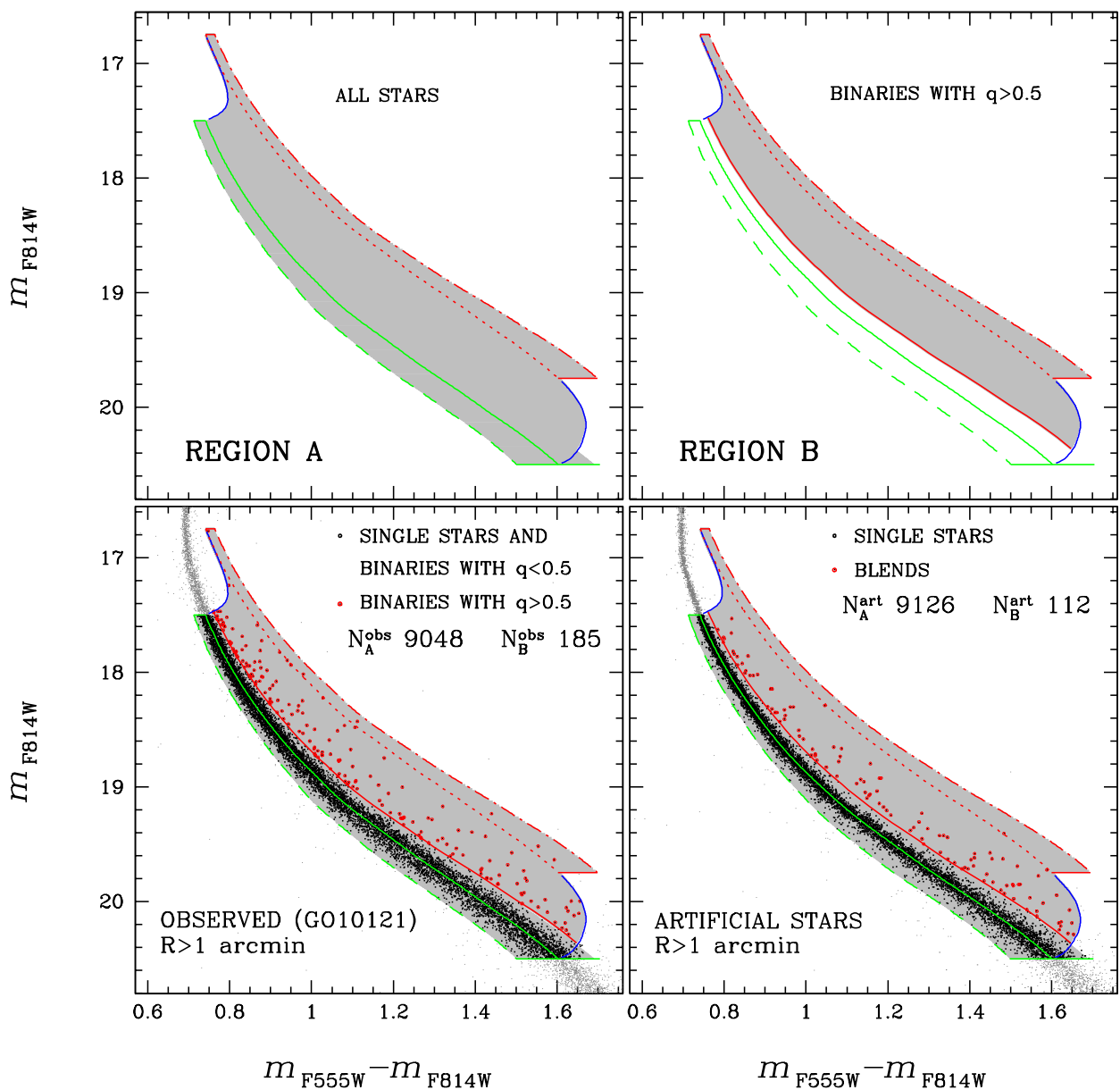}
\caption{ The upper half of the figure shows the regions A and B that
  are described in the text, delineating respectively the region of all
  cluster stars and the sub-region of binaries with mass ratio greater
  than 0.5.  In the lower half of the figure the stars are shown, with
  somewhat larger symbols for the stars in the binary
  region.  At left are the observed stars, and at right the artificial
  stars described in the text.}
\label{procedure}
\end{figure*}

In Fig.\ \ref{simulation} we compare each of the simulated CMDs with our
observed one, using the sample with the best photometry, namely the part
of the GO-10121 ACS field that has $r > 1$ arcmin.  A cursory comparison
between observed and simulated CMDs suggests that the binary fraction is
small, but we make a more accurate measure of it by using a procedure
that is illustrated in Figure \ref{procedure}.  The upper half of the
figure shows two regions of the CMD, A (upper left) and B (upper right).
Region A is drawn so as to contain all the stars that we can consider to
be cluster members.  The green continuous line is the main-sequence
ridge line, drawn by the method that we described in Sect.\ \ref{data}.
To include stars moved to the blue by measuring error, we extend region
A as far as the green dashed line, which is displaced to the blue from
the MSRL by three times the rms measuring error of the colors
(calculated as a function of $m_{\rm F814W}$).  The red dotted line is
the locus of binaries whose components have equal mass; we set the other
limit of region A by drawing the red dot-dash line, displaced this time
to the red from the dotted line by three times the rms color error.

The upper-right panel of Fig.\ \ref{procedure} shows Region B, which is
chosen in such a way that we can be reasonably sure that all stars in it
are binaries.  It commences at the locus of binaries with ratio of mass
of the secondary to that of the primary, $q=m_2/m_1=0.5$, marked by the
continuous red line, which coincidentally falls three color-error sigmas
to the red of the MSRL, and it ends at the dashed red line, which is
again the $3\sigma$ redward bound of equal-mass binaries, just as in the
upper-left panel.

The lower half shows how the stars fall with respect to these two
regions.  On the left are the observed stars; there are 9048 of them in
region A, but only 185 of these fall in region B.  On the right are
artificial stars, which we chose with random positions and with random
magnitudes but MSRL colors.  Of the 9126 ASs, 112 fell in region B.
These stars are too far from the MSRL to be explained by measuring
error; the real explanation must be that in those 112 cases two stars
fell at positions so close together that a pair of stars has blended
into a single image, which would simulate a binary.  If we
adjust for the sample size (9048 instead of 9126), then 111 apparent
binaries among the real stars are likely to be explainable as blends,
leaving only 74 binaries.

Finally, before we deduce a binary fraction, one more correction is
needed:\ for field stars.  To estimate the number of these, we used a
program devised by Girardi et al.\ (2005), which uses a Galactic model
to predict star numbers in any given direction.  For a field the size of
ours in the direction of NGC 6752, the program predicts 19 stars in
region A, 14 of which fall within region B.  We are thus left with an
estimate of 74 $-$ 14 = 60 binaries among 9048 $-$ 19 = 9029 total
stars, or 0.7 \%.  A similar procedure, applied to the central PC field
of the original RB97 images, found a binary fraction of 2.6 \%.

%

\subsection{The intrinsic broadening of the main sequence}

\begin{figure}[ht!]
    \epsscale{.7}
    \plotone{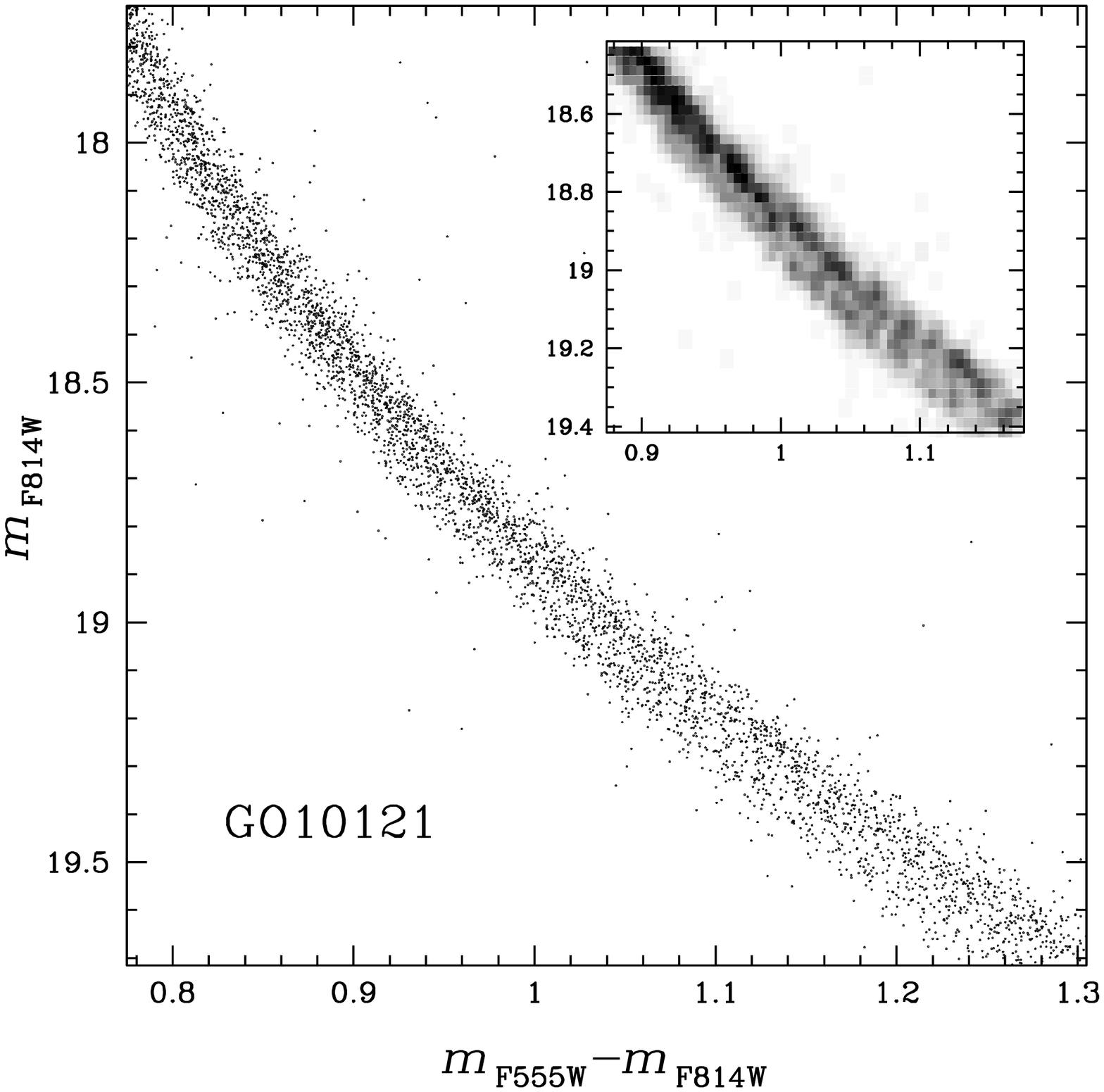}
    \caption{ CMD of NGC 6752 from GO-10121 ACS/WFC data. The inset
        shows a Hess diagram for the MS region with $18.4<
        m_{\rm F814W}<19.4$.   }
    \label{MSGO10121}
    \end{figure}

Satisfied that we are not being led astray by a serious admixture of
binaries, we now take up the question of an intrinsic breadth in the
main sequence.  As might be expected, our best source of information
comes from our best data set, that from GO-10121.  Figure
\ref{MSGO10121} shows our $m_{\rm F814W}$ vs. $m_{\rm F555W}-m_{\rm
F814W}$ CMD from this data set.  We have applied the corrections for
spatial variations of the zero point of colors, as described in
Section~\ref{data}, and have excluded stars less than 1 arcmin from the
cluster center, so as to avoid the most crowded regions.
(Note that this is a post-core-collapse cluster.)

A visual inspection of this CMD immediately suggests that NGC 6752 has a
broad MS.  The Hess diagram in the inset reinforces this impression, and
even suggests that the cluster could have a second (but less populated)
MS, on the blue side of the main MS, in close analogy with the multiple
MSs that have been observed in $\omega$ Cen and in NGC 2808.  We will
not pursue this question further, but hope that future observations will
clarify it.

We also note at this point an additional piece of evidence that the
broadening that we see is unlikely to be due to binaries.
In fact, we already showed that the fraction of binaries with mass ratio
$q>0.5$ is smaller than 3\% of the cluster population, even in the core.
In order to reproduce the broadening of the MS shown in Fig.\ 4 with
binaries, we would have to make the outlandish assumption that two
thirds of the stars in NGC 6752 are in binary systems with mass ratios
$q<0.2$.)

%

Putting aside the question of whether the main sequence is split,
and having excluded the possible contribution of binaries,
we now return to the basic question of whether it is broader than would be
expected from photometric measuring error alone.  As was shown in the
study by Anderson et al.\ (2009) of MS broadening in 47 Tuc, a very
effective way of testing for a true broadening is to divide the images
of the same field of stars into two independent sets.  If the broadening
is intrinsic, the stars that have redder (or bluer) colors in the
results from one half of the images will have redder (or bluer) colors in the
results from the other half.  But if the broadening is due only to
measuring error, a star that is redder in the first half will
have an equal chance of being redder or bluer in the second half, and
the bluer stars of the first half are equally likely to be red or blue
in the second half.  We have therefore drawn a median line in panel (a),
and used red and blue colors for the stars on either side of the line,
and in panel (b) we have kept for each star the same color that it had
in panel (a).

It is quite evident that the colors of stars are maintained very well
from panel (a) to panel (b); photometric error has led to only a small
mismatch.  To emphasize this fact, we show in panel (c) the correlation
between the colors of each star in the two halves of the data. Because
the measuring errors increase at fainter magnitudes it is clearer to
show separate plots for four successive intervals of magnitude.  If a
star was measured as redder than the MSRL in the first half of the data,
then in the great majority of cases it is also measured as redder in the
second half of the data.  This is the mark of a true spread in color.

For further emphasis we show in panel (d) the color distribution of the
straightened MS as measured from the {\it whole} set of images.  The
measuring errors of these colors are of course smaller than those of
either half of the data set, by a factor of $\sqrt{2}$, and now there is
even an intriguing hint again of the split that we suspected in Fig.\
\ref{MSGO10121}.

Finally, in panel (e) we show the distribution of half the difference
between the colors measured separately from each half of the images,
which is a statistical estimator of the errors of the colors in panel
(d).  In these last two panels we have written into each magnitude
interval an estimate of the sigma of the spread, as deduced from its
interquartile separation.

    \begin{figure}[ht!]
    \epsscale{.9}
    \plotone{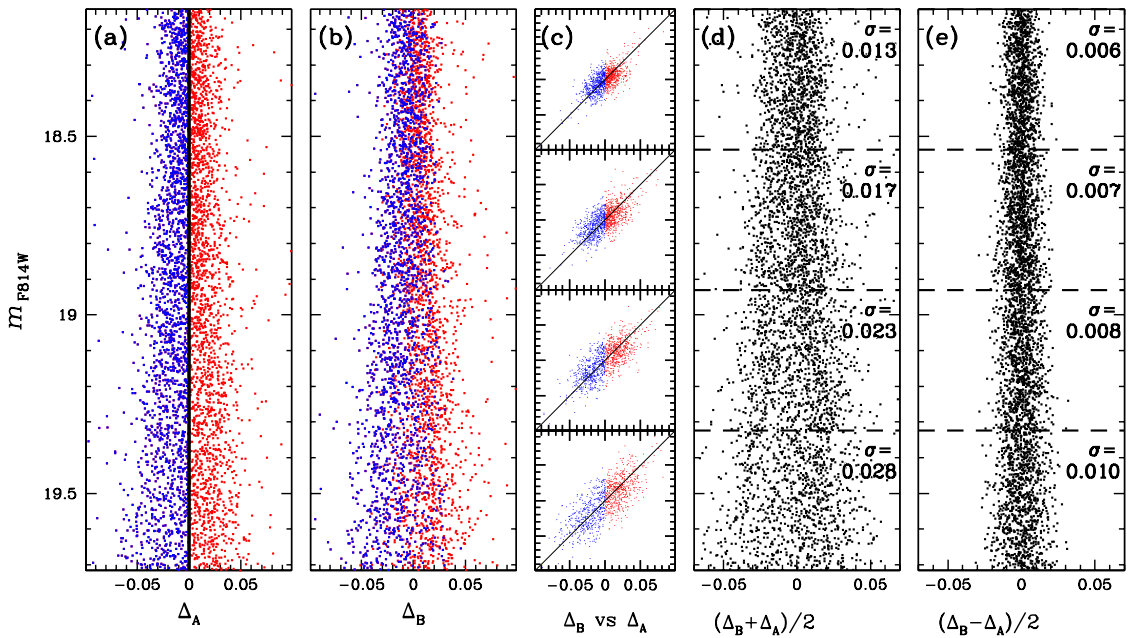}
    \caption{ (a) Straightened MSs for the first half of the GO-10121
      data.  Stars with color residuals on the blue and red sides of the
      black vertical line are colored blue and red, respectively. (b)
      Straightened MSs for the second half of the data.  Each star has
      the color that was assigned to it in the previous panel.  (c)
      Correlations of the two color residuals, in four magnitude
      intervals.  (d) Means of color residuals.  (e) Estimates of color
      errors.}
    \label{2halves}
    \end{figure}
%

An additional confirmation of the presence of an intrinsically broadened
MS in NGC 6752 comes from comparing each of the other data sets, on the
one hand, with the entire GO-10121 data set on the other hand.  Although
none of the other data sets is as good as that from GO-10121, this
course has the advantage that the exposures in each data set were
taken at different pointings and orientations, so that
a star falls on a different part of the detector in each of the two data
sets.

\begin{figure*}
   \epsscale{1.0}
   \plotone{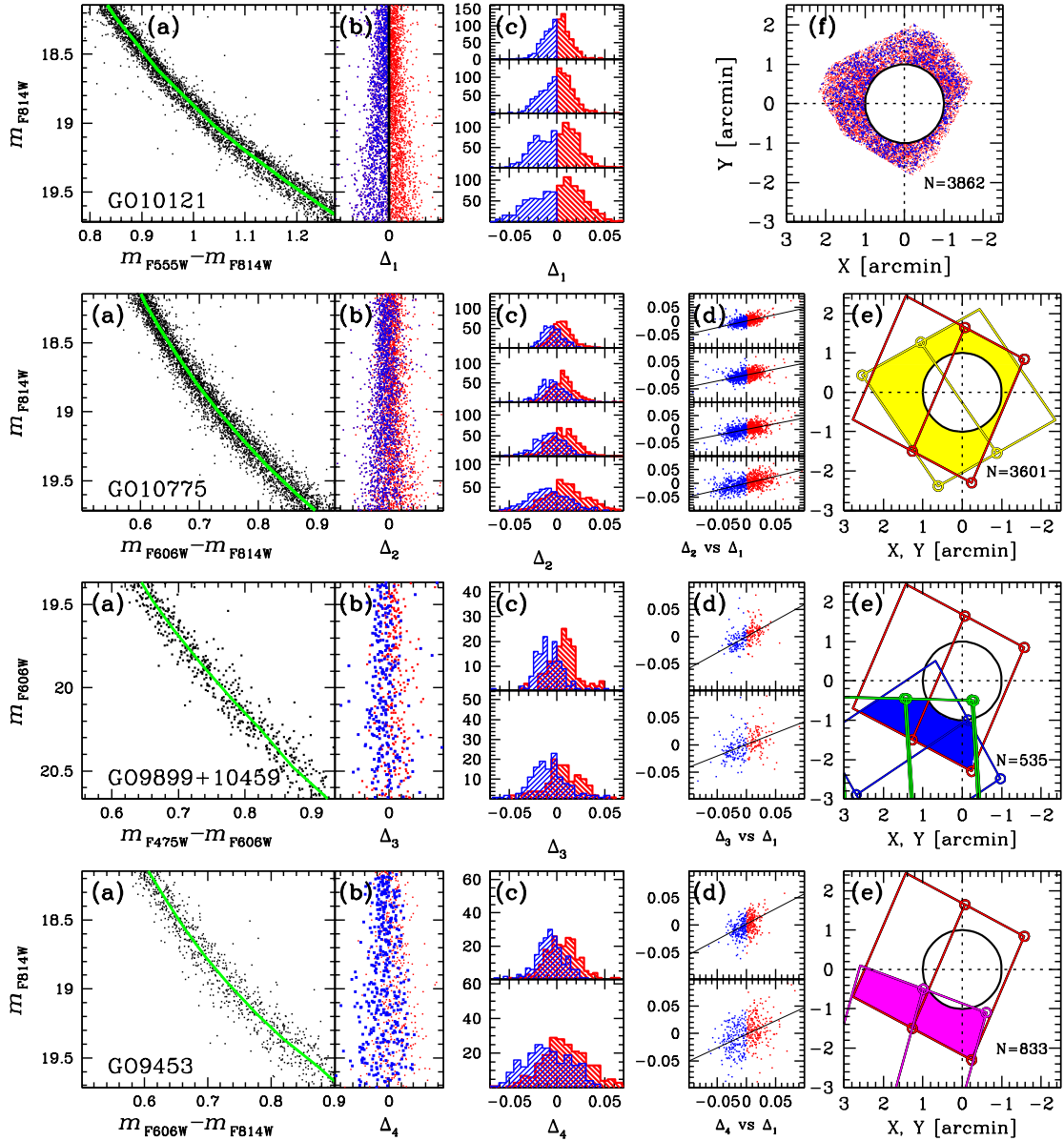}
   \caption{(a) CMDs of the MS stars, with MSRLs shown in green. (b)
     Straightened CMDs, with each star shown in the color that it was
     assigned in the straightened CMD of the GO-10121 stars.  (c)
     Histograms of the color residuals $\Delta_i$ of the stars, in
     magnitude intervals within each data set.  Shading identifies the
     stars that were marked red or blue in the top row.  Bi-colored
     cross-hatched shading is the overlap region between the blue and
     red histograms.  (d) Correlations of color residuals of each data
     set with those of the GO-10121 stars; in each, the best-fitting
     line is superimposed.  (e) Footprints of the data sets.  The stars
     used in that row are the ones from the colored area.  (f) Spatial
     distribution of the blue and red stars in the GO-10121 data set.}
   \label{4CMD}
   \end{figure*}

In Fig.~\ref{4CMD} we compare the main-sequence CMD from GO-10121 with
CMDs from three other independent ACS data sets.  Each row of panels
shows our results from one data set.
In the left panels we show the CMD, in two forms.  Part (a) is the
normal CMD; each green line is the MSRL, drawn by eye.
Again, the red/blue coding of the individual stars is defined in the
first straightened CMD, and maintained in the succeeding ones.
Part (b) shows the straightened sequence, which is the result of
subtracting from the color of each star the color of the MSRL at the
magnitude of the star; this process makes the MSRL vertical.  We have
assigned to each star in the top diagram a red or blue color code
according to whether it is on the red or blue side of the MSRL [marked
by a black dividing line in the top panel of (b)], and the star keeps
this color in all remaining straightened diagrams in part (b), and in
all of parts (c), (d), and (f).  The color displacements from the MSRL
are labeled $\Delta_i$, where the subscript $i$ goes with the row in the
figure.

This is analogous to what we showed in Fig.\ \ref{2halves}, except that
there we were comparing two halves of the GO-10121 data set with each
other, whereas here we start with the entire GO-10121 data set, and
therefore base our red-blue distinction, in panel (b) of the top row, on
a reference sample that is twice as large as the one with which we began
in the  previous figure.  In the present figure the uncertainties in
each of the comparisons shown in the remaining rows are almost
completely due to
the second data set in the comparison, which is never as strong as that
of the GO-10121 data set.  What we show here, however, is independent
confirmation, from images taken at other pointings and orientations, of
what we found from our best data set.

The panels labeled (c) show the histograms of the distribution of color
residuals ($\Delta_i$) for the stars to which the red and blue symbols
were assigned; they are divided into magnitude intervals,
in accordance with the magnitude scales
at the left edges of panels (a).  In the top row
the separation in panel (c) is of course perfect, because these are the
colors that we defined in the preceding panel.  In the other rows
the separation is less sharp, but is still strongly evident.  Table 2
gives the median values of the residuals $\Delta_i$ for the bluer stars
($\langle\Delta_i\rangle_{\rm b}$), for the redder stars
($\langle\Delta_i\rangle_{\rm r}$), and for the entire sample
($\langle\Delta_i\rangle_{\rm ALL}$).  The intrinsic color spread of the
MS is confirmed by the fact that for each of the other data sets, and in
all magnitude intervals within each of them, the residuals of the stars
that were
marked blue in the GO-10121 data set are significantly bluer than those
that were marked red.

In the fourth column of panels, labeled (d), we plot $\Delta_{i}$
against $\Delta_1$, to further show the consistency and significance of
the color spread.  The solid line is the best-fitting least-squares
straight line.  Horizontally the separation is of course perfect, since
the abscissa is $\Delta_1$; it is in the slope that we see the
correlation.  Because the different data sets have different color
baselines, we do not expect this line to have a slope of unity, but a
positive correlation indicates an intrinsic spread.

Panels (e) show the footprints of the data sets used in this analysis.
The color coding is the same as in Fig.~\ref{footprint}.  The black
circle indicates the region within 1 arcmin of the cluster center.  Only
stars in the colored areas were used in the present analysis.

\begin{figure}
   \epsscale{.7}
   \plotone{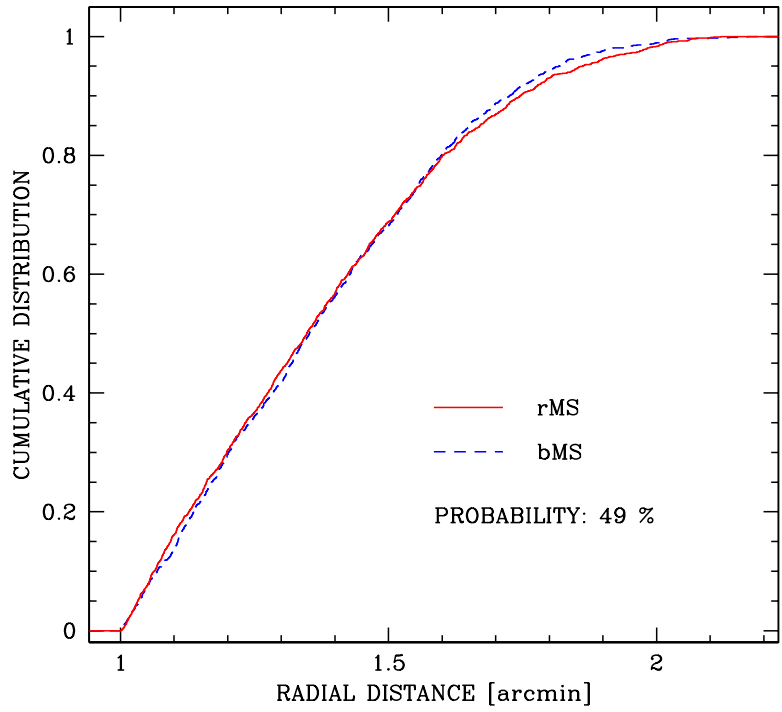}
   \caption{ Cumulative radial distributions of the bMS and rMS
   stars.} 
   \label{KSMS}
   \end{figure}

Finally, panel (f), at the upper right, shows the spatial distribution
of the stars of the GO-10121 data set, with the colors that they were
assigned in panel (b) of the top row.
 No difference in radial distribution is evident.  To test this
  better, in Figure \ref{KSMS} we show a Kolmogorov-Smirnov test; in
  random samplings from the same distribution a difference this large
  would occur 49 \% of the time.

%
%
%

\begin{table}[!ht]
\center
\scriptsize {
\begin{tabular}{cccccccc}
\hline
\hline
 CMD & mag.\ interval &  $\langle\Delta_i\rangle_{\rm b}$  & $N_{\rm b}$ &
$\langle\Delta_i\rangle_{\rm r}$  & $N_{\rm r}$ &
 $\langle\Delta_i\rangle_{\rm ALL}$
& $N_{\rm ALL}$ \\
\hline
1 & 18.1-18.5   & $-$0.008$\pm$0.001 & 480 & 0.008$\pm$0.001 & 501 & 0.000$\pm$0.001 & 981 \\
1 & 18.5-18.9   & $-$0.011$\pm$0.001 & 489 & 0.010$\pm$0.001 & 521 & 0.001$\pm$0.001 &1010 \\
1 & 18.9-19.3   & $-$0.017$\pm$0.001 & 465 & 0.014$\pm$0.001 & 481 & 0.001$\pm$0.001 & 946 \\
1 & 19.3-19.7   & $-$0.020$\pm$0.001 & 460 & 0.018$\pm$0.001 & 465 & 0.000$\pm$0.001 & 925 \\
\hline
2 & 18.1-18.5   & $-$0.005$\pm$0.001 & 478 & 0.005$\pm$0.001 & 498 & 0.000$\pm$0.001 & 976 \\
2 & 18.5-18.9   & $-$0.005$\pm$0.001 & 501 & 0.006$\pm$0.001 & 514 & 0.001$\pm$0.001 &1015 \\
2 & 18.9-19.3   & $-$0.006$\pm$0.001 & 480 & 0.006$\pm$0.001 & 478 & 0.000$\pm$0.001 & 958 \\
2 & 19.3-19.7   & $-$0.009$\pm$0.001 & 453 & 0.009$\pm$0.001 & 461 & 0.002$\pm$0.001 & 914 \\
\hline
3 & 19.4-20.0   & $-$0.006$\pm$0.001 & 135 & 0.008$\pm$0.001 & 138 & 0.001$\pm$0.001 & 273 \\
3 & 20.0-20.7   & $-$0.009$\pm$0.002 & 125 & 0.008$\pm$0.002 & 136 & 0.000$\pm$0.001 & 261 \\
\hline
4 & 18.1-18.9   & $-$0.006$\pm$0.001 & 199 & 0.006$\pm$0.001 & 207 & 0.000$\pm$0.001 & 406 \\
4 & 18.9-19.7   & $-$0.011$\pm$0.002 & 205 & 0.008$\pm$0.001 & 222 & 0.000$\pm$0.001 & 427 \\
\hline \hline
\end{tabular}
}

\label{tabMS}
\caption{
Median color residuals for blue MS stars
  ($\langle\Delta_i\rangle_{\rm b}$), red MS stars
  ($\langle\Delta_i\rangle_{\rm r}$), and for the whole sample of MS
  stars ($\langle\Delta_i\rangle_{\rm ALL}$), for each data set, in the
  magnitude intervals into which we have divided them. }
\end{table}

\newpage
\section{The subgiant branch}
\label{SGB}
%

\begin{figure*}
   \epsscale{1.0}
   \plotone{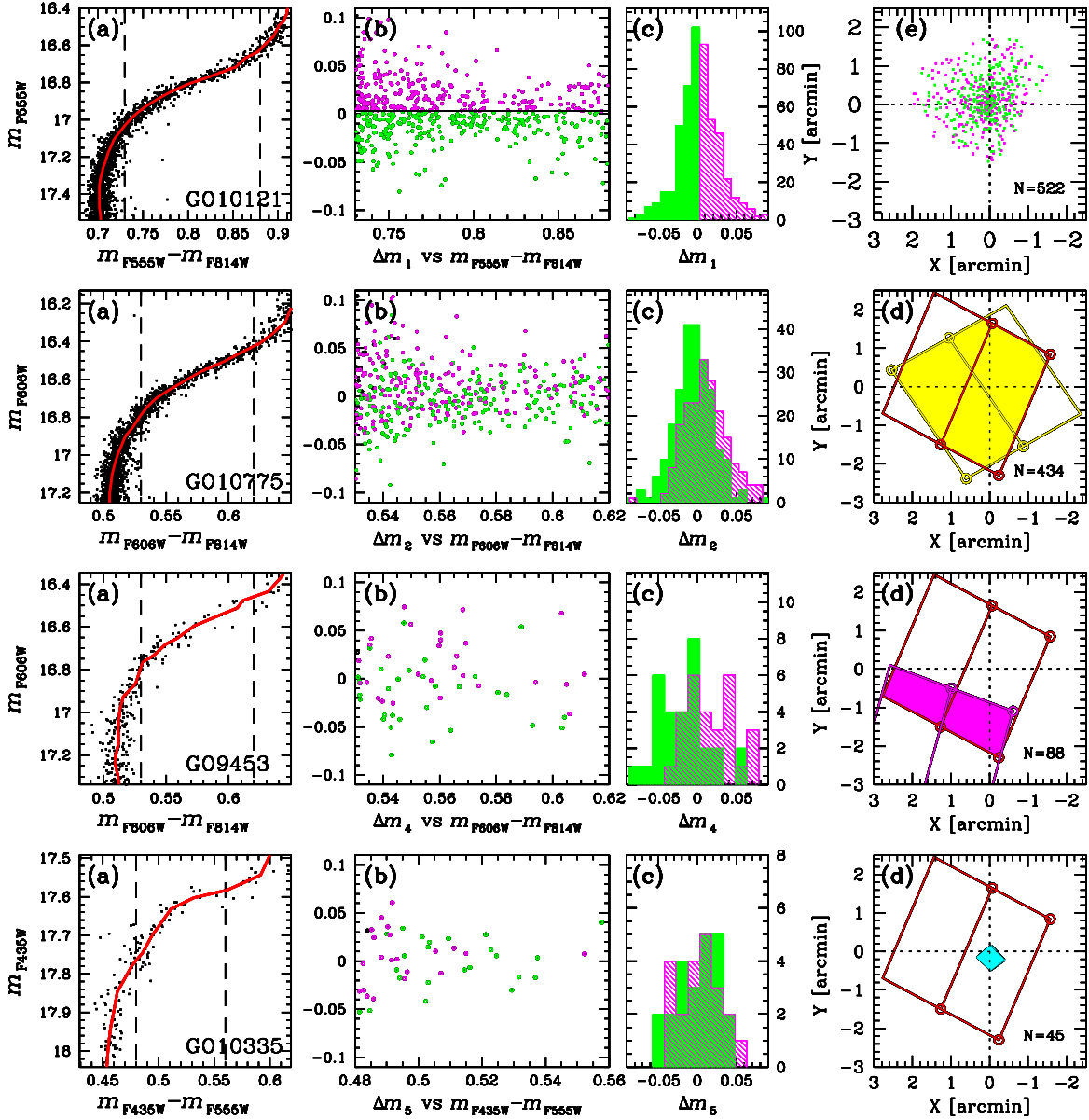}
   \caption{(a) Zoom of the CMDs around the SGB.  (b) Straightened SGBs,
     with the samples of fSGB, and bSGB stars (faint, and
     bright SGB) colored green, and magenta, respectively.
     In the top row, the stars are from the
     GO-10121 CMD, in the other rows, from the colored area in
     panels (d).  (c) Distributions of the magnitude residuals $\Delta m_{i}$.
     (d) Footprints of the data sets.  The stars used in this analysis
     come from the colored area.  (e) (at upper right): Spatial
     distributions of the fSGB and bSGB stars (in green, and
     magenta, respectively).}
   \label{4SGB}
   \end{figure*}

In this section we look for any evidence of multiple populations among
the SGB stars of NGC 6752.  The procedure that we used is illustrated in
Fig.~\ref{4SGB}.  In the (a) panels we show the independent CMDs from
data sets 1, 2, 4 and 5 (Sect.\ \ref{data}), zoomed around the region of
the SGB.  Here we
do include the central part of the cluster, because at this magnitude
level crowding is not a significant photometric problem.  A visual
inspection at these CMDs does not reveal any significant SGB split, or
even a broadening, either in magnitude or in color.  There might be a
few outliers below the SGB, but it is difficult to assess whether they
could represent a different population, as, for instance, in NGC 1851
(Milone et al.\ 2008) or M22 (Marino et al.\ 2009),
or else blends/poor photometry.

In order to provide an upper limit for the SGB spread, we drew in, by
hand, the SGB ridge-lines (SGB-RLs) that are shown as red solid lines in
panels (a) of Fig.~\ref{4SGB}, and
computed magnitude residuals $\Delta m_{i}$ by subtracting
from the magnitude of each star
the magnitude of the SGB-RL at the color of the star. The residuals
(note that here they are magnitude residuals, rather than color
residuals, as in the earlier figures) are plotted against color in
panels (b), and histograms of the $\Delta m_{i}$ distributions are shown
in panels (c), while panels (d) show the footprints of the data sets,
using the same color coding as in Figs.\ \ref{footprint} and \ref{4CMD}.
Panel (e), at the top right, shows the spatial distributions of the
three magnitude subsets that we defined in panel (b) of that same row.

Once again, the CMD from the GO-10121 images has the best photometric
precision, and we used it to define, in the (a) panel for that data set,
 two subsamples of stars brighter (bSGB) and fainter (fSGB) than the SGB-RL.
We chose magenta and green for bSGB and fSGB stars respectively.
In all the other diagrams we use the same 
color code for each star that it has in the GO-10121 data set.  In the
plots for each of the other data sets the median $\Delta m$'s of the
bSGB and the fSGB stars are, respectively, only marginally brighter and
fainter than the median $\Delta m$ of all SGB stars.  The individual
values are given in Table \ref{tabSGB}, and confirm that no significant
intrinsic spread of the SGB of NGC 6752 is detectable in our data.  Any
intrinsic spread must be smaller than a few hundredths of a
magnitude.

\begin{table}[!ht]
\center
\scriptsize {
\begin{tabular}{cccccccccc}
\hline
\hline
 CMD &  $\langle\Delta m_{i}\rangle_{\rm bSGB}$ & $N_{\rm bSGB}$
     &  $\langle\Delta m_{i}\rangle_{\rm fSGB}$ & $N_{\rm fSGB}$
     &  $\langle\Delta m_{i}\rangle_{\rm ALL}$  & $N_{\rm ALL}$  \\
\hline
1 &    0.018$\pm$0.001 & 249 & $-$0.011$\pm$0.001 &273 &      0.001$\pm$0.001 & 522 \\
2 &    0.010$\pm$0.002 & 197 & $-$0.004$\pm$0.002 &233 &      0.002$\pm$0.001 & 430 \\
3 &    0.012$\pm$0.006 &  28 & $-$0.014$\pm$0.005 & 30 &   $-$0.003$\pm$0.005 &  58 \\
5 &    0.004$\pm$0.006 &  21 &    0.003$\pm$0.006 & 23 &      0.004$\pm$0.004 & 44  \\
\hline
\end{tabular}
}
\label{tabSGB}
\caption{Average magnitude residuals for bright SGB stars
($\langle\Delta m_{i}\rangle_{\rm bSGB}$), 
faint SGB stars
($\langle\Delta m_{i}\rangle_{\rm fSGB}$), and for all the SGB stars
($\langle\Delta_i\rangle_{\rm ALL}$) in data sets 1, 2, 3, and 5.}
\end{table}

Note that this result really comes only from the GO-10121 and GO-10775
data sets.  The other data sets have too few stars to contribute
anything significant; we show them only for completeness, and to verify
that they do not show anything that would contradict our conclusion.

\begin{figure}
   \epsscale{.7}
   \plotone{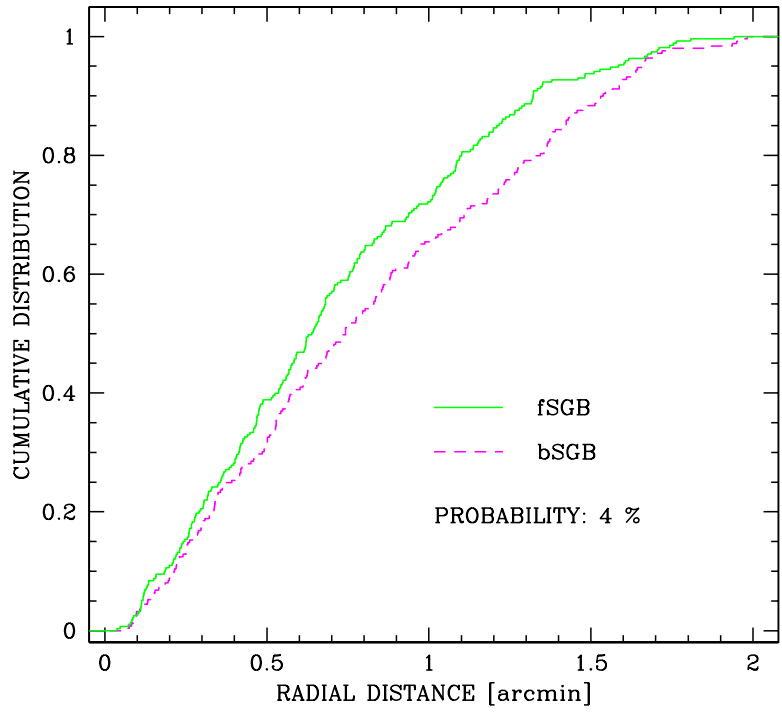}
   \caption{ Cumulative radial distributions of the bSGB and fSGB stars.}
   \label{KMSGB}
   \end{figure}

 As for the radial distributions, Figure \ref{KMSGB} compares them
for bSGB and fSGB stars.  Although the difference would appear to be
large, the Kolmogorov-Smirnov statistic indicates that in random
samplings from the same distribution a difference this large would occur
4 \% of the time, suggesting that a real difference is possible, but far
from conclusive.

%

%

\newpage
\section{Na and O abundances along the broad red-giant branch}
\label{RGB}
%

Photometric evidence for a broadening of the RGB of NGC 6752 was found
by Grundahl et al.\ (2002) and Yong et al.\ (2008). Both papers found that the
Str\"omgren
photometric index $c_1$ correlates with the nitrogen abundance, and
concluded that the observed scatter in this index is due to variations
in NH band strength, in stars both brighter and fainter than the RGB
bump.

\begin{figure}
   \epsscale{1.0}
   \plotone{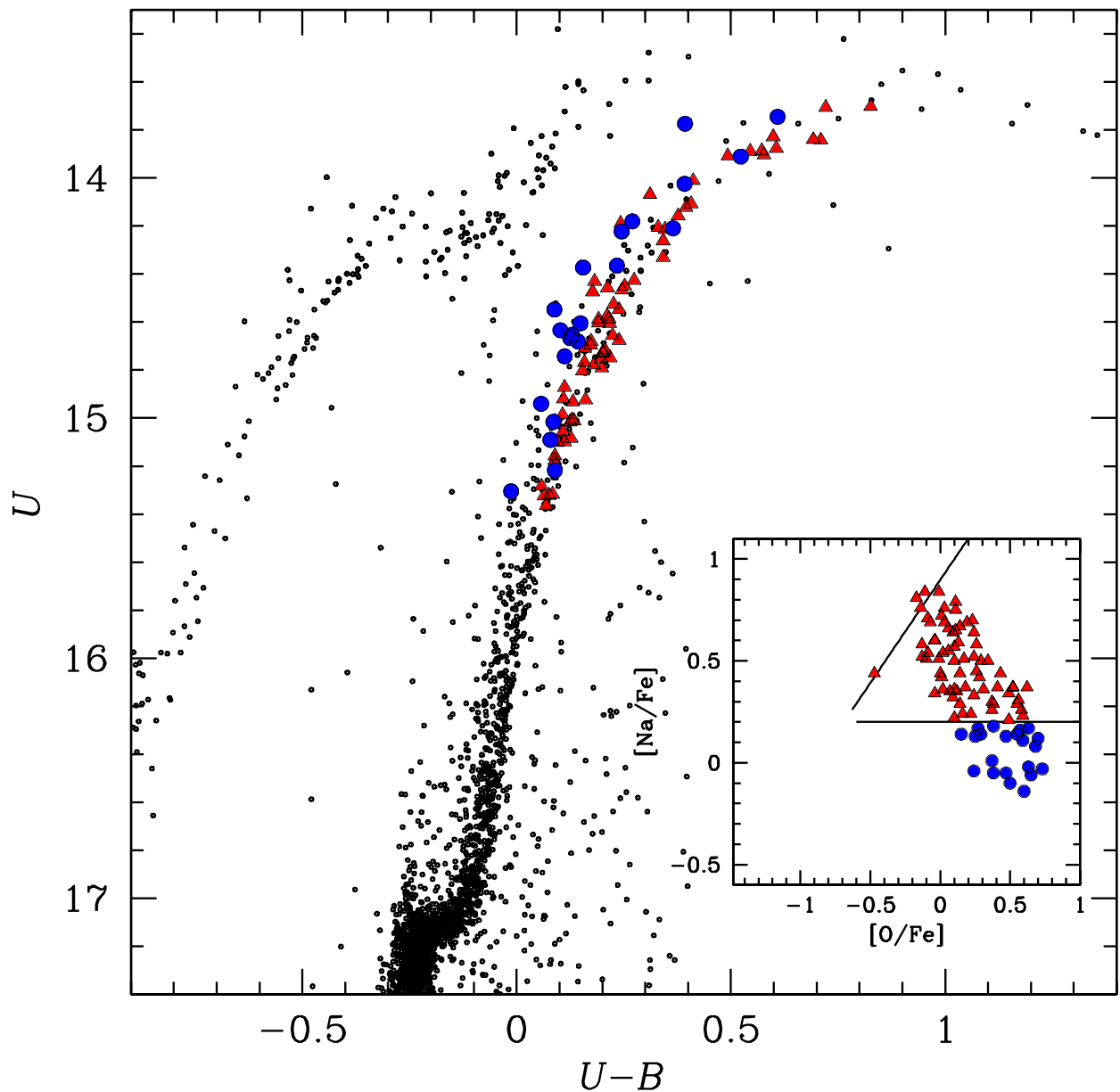}
  \caption{$U$ vs.\ $(U-B)$ CMD from Wide Field Imager photometry,
     as obtained by Momany et al.\ (2002).
     The Na-O anti-correlation
     for the stars measured by Carretta et al.\ (2007) is shown in the
     inset, where the solid lines delimit the primordial, intermediate,
     and extreme populations, as defined in Carretta et al.\
     The stars of the primordial population are represented by blue
     circles, while red triangles are the intermediate and extreme
     groups.}
   \label{figRGB}
   \end{figure}

A broadened RGB is also visible in a $U$ vs.\ $U-B$ CMD of NGC 6752,
derived from ground-based photometry by Momany et al.\ (2002) with ESO's
Wide Field Imager, and shown in Fig.\ \ref{figRGB}.  The broadening
extends from
the base of the RGB to the RGB tip, and looks extremely similar to the
broadening of the RGB observed in M4, and associated by Marino et al.\
(2008) with the presence of two stellar populations with different
ratios in their [Na/Fe] and [O/Fe] abundances, and different CN band
strengths.  The two groups of stars studied by Marino et al.\ (2008) in
M4 lie in two different regions of the RGB. One group, which consists of
Na-rich, CN-strong stars, occupies a narrow sequence on the red side of
the RGB, while the other group, made up of Na-poor and CN-weak stars,
has a broader spread on the blue side of the RGB.  (See Marino et al.\
2008 for more details on the abundances of RGB stars in M4.)

The Na-O anticorrelation in NGC 6752 has been extensively studied by
Gratton et al.\ (2001), and Carretta et al.\ (2005, 2007,
2009).
In particular, both Gratton et al.\ (2001) and the 2005 Carretta et
al.\ paper
also found the Na-O anti-correlation in unevolved and barely evolved
stars.  Carretta et al.\ (2007,
2009) present an analysis of the largest
sample of globular cluster RGB stars available in the literature.  In Fig.\ \ref{figRGB}
we identify stars from Carretta et al.\ (2007) in our $U$ vs.\ $U-B$
CMD, by using the same symbols as in the inset, which shows the Na-O
anticorrelation that they found.  Here, following the criteria
defined in Carretta et al.\
(2009),
we isolate two subsamples:\ $i)$ the
primordial population, located in the Na-poor/O-rich region (marked with
blue symbols), and $ii)$ their extreme and intermediate
components, which are Na-rich/O-poor.  (We have lumped these two groups
together as red symbols).

The same two groups of stars are located in different regions of the RGB
in the $U$ vs.\ $U-B$ CMD, in close analogy with the results by Marino
et al.\ (2008) for M4.  Most of the Na-poor (primordial) stars are
distributed on a bluer and broader sequence around the RGB, while the
Na-rich stars tend to be distributed on a narrower sequence on the red
part of the RGB.
This stellar distribution continues well below the HB
level.
The bimodal color distribution also extends well
below the RGB-bump, suggesting that it is not a consequence of the
mixing of the stellar interiors.  The MS of NGC 6752 is intrinsically
broadened, and possibly split.
Also, the Na-O anti-correlation is more extended in NGC 6752 than in M4.

%
%

\newpage
\section{Discussion}
%

The CMDs presented in this paper demonstrate that NGC 6752 hosts a
stellar population that has complexities similar to those
identified in other GCs investigated so far (see Piotto 2009 for a
recent review).  This cluster exhibits a broadened MS, with some
evidence of a split, while the SGB does not show any significant
vertical spread, even at the level of a few hundredths of a magnitude.
The MS broadening is intrinsic, and not due to binaries or photometric
errors.  The RGB is significantly broadened both in the Str\"omgren
$c_1$ index (Yong et al.\ 2008) and in the $U-B$ color.  Stars that are
O-poor and Na-rich mainly populate a sequence on the red RGB portion of
the $U$ vs.\ $U-B$ CMD, while O-rich/Na-poor stars are preferentially
located on the blue side of the RGB, with a broader color distribution.

It is interesting to look for similarities between NGC 6752 and the
other clusters with photometric evidence of multiple stellar
populations.  The only two other clusters that are known to have
multiple MSs (i.e., with clearly separated branches, rather
than just a broadening) are $\omega$~Cen and NGC 2808.  Starting from
the CMD morphologies and the information we have on their chemical
abundance patterns, and by analogy with the current interpretation of the
multiple-MS phenomenon, we can hypothesize that the stars on the
blue side of the MS of NGC 6752 could belong to a second stellar
generation, born from material polluted by a previous population of
stars, and He-enriched.  We note that a large amount of He has also been
invoked to interpret the blue HB tails which are present in many
clusters (D'Antona et al.\ 2005).  Interestingly, $\omega$ Centauri, NGC
2808, M54, and NGC 6752 all have such extended blue HBs.

In an increasing number of clusters, including NGC 1851, M22, and
NGC 6388, the presence of multiple stellar populations is inferred from
the split SGB, but no evidence of a split or spread-out MS has been
found so far in these objects.  (Some evidence of a split in the SGB has
been found in 47 Tuc too, by Anderson et al.\ 2009, though that case
lacks the clear separation shown in the other clusters that we have
mentioned.)  The split SGB of NGC 1851 has been interpreted as
coming from two stellar populations with large differences in the
total
C+N+O abundance (Cassisi et al.\ 2008, Ventura et al.\ 2009).  Indeed,
large variations in the total C+N+O content among four bright RGB stars
of NGC 1851 have been reported by Yong et al.\ (2009).  In NGC 1851 and
M22 the two SGBs may be related to the two groups of stars with a
large difference in the abundances of $s$-process elements ($\sim$ 0.4
dex), identified by Yong \& Grundahl (2008) and Marino et al.\ (2009),
respectively.

In NGC 6752 the overall sum C+N+O is almost constant (Carretta et al.\
2007).  This cluster presents a
large
star-to-star abundance variation in N
(1.95 dex, Yong et al.\ 2008), and the amplitude of the abundance
variations for $s$-process elements is small ($<$0.2 dex, Yong et al.\
2008).

If the polluting matter is identified with the envelopes of asymptotic
giant branch stars (Ventura et al.\ 2001, 2002), these AGB stars
should be very massive in order to experience the second dredge-up,
which brings a large amount of fresh He to the envelope, and the hot
bottom burning process, which converts C into N.  Such massive AGB stars
should not live long enough, however, for a sufficient number of them to
experience a third dredge-up and significantly increase the total C+N+O
abundance in the envelope (Renzini 2008).  As an alternative, polluted
material could come from the envelopes of fast-rotating stars during the
H-burning phase; in this case too, we expect a He-enriched second
generation without any C+N+O enhancement at all (Decressin et al.\
2007).

At this stage, observational insight into such processes could come from
a detailed spectroscopic analysis of stars on the blue and red sides of
the MS and of the RGB.  In addition, the new {\sl WFC3} camera, which is
sensitive to both the UV and IR spectral regions, should give a wide
color baseline, offering a unique opportunity to better distinguish the
signatures of multiple stellar populations in the CMD.  Such a study
would allow not only a better characterization of the complex case of
NGC 6752, but also a more consistent interpretation of the observations
of multiple stellar populations in GCs in general.


\bigskip
{\it Acknowledgements} We are very thankful to Raffaele Gratton and
Eugenio Carretta for generously sharing with us their spectroscopic
catalogs of Na and O abundances in stars, in electronic format, and for
the many fruitful discussions.  We also thank Antonio Sollima for 
useful discussion. APM, AFM, and GP acknowledge support by
PRIN2007 (prot.\ n.\ 20075TP5K9), and by ASI under the program ASI-INAF
I/016/07/0.  IRK and JA acknowledge support by STScI grants
GO9899
GO-10922, and
GO-11233. LM acknowledges support by the STScI {\it 2009 Space Astronomy Summer
Program}.


\begin{thebibliography}{}

\bibitem[Anderson \& King (1999)]{anderson99}
	Anderson, J., \& King, I.\ R. 1999, \pasp, 111, 1095

\bibitem[Anderson \& King (2000)]{anderson00}
	Anderson, J., \& King, I.\ R. 2000, \pasp, 112, 1360

\bibitem[Anderson \& King (2003)]{anderson03}
Anderson, J., \& King, I.\ R. 2003, \pasp, 115, 113

\bibitem[Anderson et al.\ (2004)]{anderson04}
	Anderson, J., \& King, I.\ R. 2004, ACS ISR 2004-15

\bibitem[Anderson \& King (2006)]{anderson06}
         Anderson, J., \& King, I.\ R. 2006, ACS ISR 2006-01

\bibitem[Anderson et al.\ (2008)]{anderson08}
	Anderson, J., et al.\ 2008, AJ, 135, 2055

\bibitem[Anderson et al.\ (2009)]{anderson09}
	Anderson, J., Piotto, G., King, I.\ R.,
 	 Bedin, L.\  R., \& Guhathakurta, P.\ 2009, ApJ, 697, L62

\bibitem[Bedin et al.\ (2001)]{bedin01} Bedin, L.\ R., Anderson, J.,
King, I.\ R., \& Piotto, G.\ 2001, ApJ, 560, L75

\bibitem[Bedin et al.\ (2004)]{bedin04}
         Bedin, L.\  R., Piotto, G., Anderson, J., Cassisi, S., King, I.\ R.,
         Momany, Y., \& Carraro, G.  2004, ApJ, 605, L125

\bibitem[Bedin et al.\ (2005)]{bedin05}
         Bedin, L.\  R., Cassisi, S., Castelli, F., Piotto, G.,
         Anderson, J., Salaris, M., Momany, Y., \& Pietrinferni, A.
         2005, \mnras, 357, 1038

\bibitem[Carretta et al.\ (2005)]{carretta05}
	Carretta, E., Gratton, R.\ G., Lucatello, S., Bragaglia, A.,
	\& Bonifacio, P. 2005, A\&A, 433, 597

\bibitem[Carretta et al.\ (2007)]{carretta07}
	Carretta, E., Bragaglia, A., Gratton, R.\ G., Lucatello, S.,
	\& Momany, Y. 2007, A\&A, 464, 927

\bibitem[Carretta et al.\ (2008)]{carretta08}
Carretta, E. et al.\ 2009, A\&A, 505, 117

\bibitem[Cassisi et al.\ (2008)]{cassisi08}
         Cassisi, S., Salaris, M., Pietrinferni, A.,
         Piotto, G., Milone, A.\ P., Bedin, L.\  R., \& Anderson, J. 2008,
         ApJ, 672, L115

\bibitem[D'Antona et al.\ (2005)]{dantona05}
	D'Antona, F., Bellazini, M., Caloi, V., Fusi
  	Pecci, F., Galleti. S., \& Rood, R.\ T. 2005, \apj, 631, 868

\bibitem[Decressin et al.\ (2007)]{decressin07}
	Decressin, T., Meynet, G., Charbonnel, C., Prantzos, N.,
	\& Ekstr\"om, S. 2007, A\&A, 464, 1029

\bibitem[Girardi et al. \ (2005)]{girardi05}
	Girardi, L., Groenewegen, M.\ A.\ T., Hatziminaoglou, E., \&
	da Costa, L. 2005, A\&A 436, 895

\bibitem[Gratton et al.\ (2001)]{gratton01}
         Gratton, R.\ G., et al.\ 2001, \aap, 369, 87

\bibitem[Gratton et al.\ (2004)]{gratton04}
         Gratton, R.\ G., Sneden, C., \& Carretta, E.  2004, ARA\&A, 42, 385

\bibitem[Grundahl et al.\ (2002)]{grundahl02}
Grundahl, F., Briley, M., Nissen, P.\ E., \& and Feltzing, S. 2002,
A\&A, 385, L14

\bibitem[Harris  (1996)]{harris96}
         Harris, W.\ E.  1996, AJ, 112, 1487 (February 2003 update)

\bibitem[Kroupa\ 2002] {kroupa02}
	Kroupa,  P.\ 2002, in ASP Conf.\ Ser.\ 285, 
	Modes of Star Formation and the Origin of Field 
	Populations, eds.\ E.\ K.  Grebel \& W.\ Brandner
	(San Francisco: ASP), p.\ 86

\bibitem[Landolt.\ (1992)]{landolt92}
	Landolt, A.\ U. 1992, \aj, 104, 372

\bibitem[Lee et al.\ (1999)]{lee99}
         Lee, Y.-W., Joo, J.-M., Sohn, Y.-J., Rey, S.-C.,
         Lee, H.-C., \& Walker, A.\ R. 1999, Nature, 402, 55

\bibitem[Mackey et al.\ (2008)]{mackey08}
	Mackey, A.\ D., Broby Nielsen, P., Ferguson, M.\ N.,
	\& Richardson, J.\ C. 2008, ApJ, 681, L17

\bibitem[Marino et al.\ (2008)]{marino08}
	Marino, A.\ F.,  Villanova,  S., Piotto,  G.,
	Milone,  A.\ P., Momany,  Y., Bedin,  L.\ R., \&
	Medling,  A.\ M. 2008, \aap, 490, 625

\bibitem[Marino et al.\ (2009)]{marino09}
	Marino, A.\ F.,	Milone,  A.\ P.,  Piotto,  G., Villanova,  S.,
	Bedin,  L.\ R., \& Renzini,  A.\ 2009, A\&A, 505, 1099

\bibitem[Milone et al.\ (2008)]{milone08}
         Milone, A.\ P., et al.\ 2008, ApJ, 673, 241

\bibitem[Milone et al.\ (2008b)]{milone08b}
	 Milone, A.\ P., Piotto, G., Bedin, L.\ R. \&
	 Sarajedini, A. 2008b, in XXI Century Challenges for
	 Stellar Evolution, Memorie della Societa Astronomica Italiana,
	 vol. 79/2, eds: S. Cassisi, M. Salaris (arXiv:0801.3177)

\bibitem[Milone et al.\ (2009)]{milone09}
         Milone, A.\ P., Bedin, L.\ R., Piotto, G., \&
         Anderson, J. 2009, A\&A, 497, 755

\bibitem[Momany et al.\ (2002)]{momany02}
	Momany, Y., Piotto, G., Recio-Blanco, A., Bedin, L.,\ R.,
	Cassisi, S., \& Bono, G.  2002 ApJ, 576, L65

\bibitem[Moretti et al.\ (2009)]{moretti09}
         Moretti, A., et al.\ 2008, A\&A, 493, 539

\bibitem[Norris 2004]{norris04}
	Norris, J.\ E.\ 2004, ApJ, 612, L25

\bibitem[Pancino et al.\ (2000)]{pancino00}
	Pancino, E., Ferraro, F.\ R., Bellazzini, M., Piotto, G., \&
        Zoccali, M. 2000, ApJ, 534, L83

\bibitem[Piotto et al.\  (2005)]{piotto05}
         Piotto, G., et al.\ 2005, \apj, 621, 777

\bibitem[Piotto et al.\  (2007)]{piotto07}
          Piotto, G., Bedin, L.\  R., Anderson, J., King, I.\  R., Cassisi, S.,
          Milone, A.\ P., Villanova, S., Pietrinferni, A., \& Renzini, A.
                  2007, ApJ, 661, L53

 \bibitem[Piotto 2009]{piotto09}
         Piotto, G., 2009, in IAU Symposium No.\ 258, the Ages of Stars,
	eds.\  E.\ E. Mamajek, D.\ R. Soderblom, \& R.\ F.\ G.\ Wyse
	(Cambridge: Cambridge Univ. Press), p.\ 233 (arXiv:0902.1422)


\bibitem[Renzini \& Buzzoni \ (1986)]{renzini86}
	Renzini, A., \& Buzzoni, A. 1986, in Spectral Evolution
	of Galaxies, eds.\
	C.\ Chiosi \& A.\ Renzini (Dordrecht:\ Reidel), ASSL, 122, p.\ 195

\bibitem[Renzini\ (2008)]{renzini08}
	Renzini, A. 2008, \mnras, 391, 354

\bibitem[Rubenstein \& Bailyn (1997)]{rubenstein97}
        Rubenstein, E.\ P., \& Bailyn, C.\ D. 1997, ApJ, 474, 701
        [RB97]

\bibitem[Sirianni et al.\ (2005)]{sirianni05}
         Sirianni, M., et al.\ 2005, \pasp, 117, 1049

\bibitem[Sollima et al.\ (2007)]{sollima07}
	Sollima, A., Beccari, G., Ferraro, F.\ R., Fusi Pecci, F., \& Sarajedini, A. 2007, \mnras, 380, 781

\bibitem[Stetson et al.\ (1994)]{stetson94}
	Stetson, P.\ B. 1994, \pasp, 106, 250

\bibitem[Valdes \ (1998)]{valdes98}
	Valdes, F.\ G. 1998, in ASP Conf.\ Ser.\ 145, Astronomical Data Analysis
	Software and Systems VII, eds.\ R. Albrecht, R.\ N.\ Hook, \& H.\ A.\
	Bushouse (San Francisco: ASP), p.\ 53

\bibitem[Ventura et al.\ (2001)]{ventura01}
	Ventura, P., D'Antona, F., Mazzitelli, I.,
	\&  Gratton, R. 2001, \apj, 550, 65

\bibitem[Ventura et al.\ (2002)]{ventura02}
	Ventura, P., D'Antona, F., \& Mazzitelli, I. 2002, A\&A 393, 215

\bibitem[Ventura et al.\ (2009)]{ventura09}
	Ventura, P., Caloi, V., D'Antona, F., Ferguson, J., Milone, A.,
	\&  Piotto, G. 2009, \mnras, in press (arXiv:0907.1765)

\bibitem[Villanova et al.\ (2007)]{villanova07}
         Villanova, S., et al.\ 2007, \apj, 663, 296

\bibitem[Young et al.\ (2008)]{yong08}
	Yong, D.,  Grundahl, F., Johnson, J.\ A.,
	\& Asplund, M. 2008, \apj, 684, 1159

\bibitem[Young \&  Grundahl\ (2008)]{yong08}
	Yong, D., \&  Grundahl, F.  2008, ApJ, 672, L39

\bibitem[Young et al.\ (2009)]{yong09}
	Yong, D.,  Grundahl, D'Antona, F., Karakas, A.\ I.,
	Lattanzio, J.\ C., \& Norris, J.\ E. 2009, \apj, 695, L72

\end{thebibliography}
\end{document}